\documentclass{ws-procs9x6}

\def\etal {{\it et al.}}

\begin{document}

\title{CPT TEST WITH (ANTI)PROTON MAGNETIC MOMENTS\\
BASED ON QUANTUM LOGIC COOLING AND READOUT} 

\author{M.\ NIEMANN,$^a$ A.-G.\ PASCHKE,$^a$ T.\ DUBIELZIG,$^a$\\ 
S.\ ULMER,$^b$ and C.\ OSPELKAUS$^{a,c,*}$}

\address{
$^a$Institute of Quantum Optics, Leibniz Universit\"at Hannover\\
Welfengarten 1, 30167 Hannover, Germany\\
$^b$RIKEN Advanced Science Institute\\
Hirosawa, Wako, Saitama 351-0198, Japan\\
$^c$Physikalisch-Technische Bundesanstalt\\
Bundesallee 100, 38116 Braunschweig, Germany\\
$^*$E-mail: christian.ospelkaus@iqo.uni-hannover.de}

\begin{abstract}

Dehmelt and VanDyck's famous 1987 measurement of the electron and positron $g$-factor is still the most precise $g$-factor comparison in the lepton sector, and a sensitive test of possible CPT violation. A complementary $g$-factor comparison between the proton and the antiproton is highly desirable to test CPT symmetry in the baryon sector. Current experiments, based on Dehmelt's continuous Stern-Gerlach effect and the double Penning-trap technique, are making rapid progress. They are, however, extremely difficult to carry out because ground state cooling using cryogenic techniques is virtually impossible for heavy baryons, and because the continous Stern-Gerlach effect scales as $\mu/m$, where $m$ is the mass of the particle and $\mu$ its magnetic moment. Both difficulties will ultimately limit the accuracy. We discuss experimental prospects of realizing an alternative approach to a $g$-factor comparison with single (anti)protons, based on quantum logic techniques proposed by Heinzen and Wineland 
and by Wineland \etal\
The basic idea is to cool, control and measure single (anti)\-protons through interaction with a well-controlled atomic ion. 
\end{abstract}

\bodymatter

\section{Classical approach}

Comparisons of $g$-factors between particles and their antiparticles are a sensitive test of CPT symmetry\cite{bluhm_testing_1997,bluhm_CPT_1998}. 
With leptons, the most sensitive tests carried out so far are based on the continuous Stern-Gerlach effect developed by Dehmelt and coworkers\cite{dehmelt_csge}. In essence, a $g$-factor measurement is realized by measuring the Larmor frequency $\omega_L=(g/2) (q/m) B$, which is a measure of the energy splitting between the two spin states at a given magnetic field $B$, and the free cyclotron frequency $\omega_C=(q/m) B$. The $g$-factor is then given by $g=2(\omega_L/\omega_C)$. Measurements of these frequencies are typically carried out in Penning traps, which feature a large, static $B$ field for radial confinement. For the electron, frequencies exceed 100\,GHz and allow cryogenic cooling of the cyclotron frequency to the ground state when in equilibrium with a dilution refrigerator. Motional frequencies can be measured using the image charges induced by the particle's motion in the trap electrodes. The well-known invariance theorem\cite{gabrielse_why_2009} relates the measured motional frequencies of the trapped particle to the free cyclotron frequency $\omega_C$. Superimposing a slight axial magnetic field curvature makes the trap frequency spin-dependent; readout of the spin state as required for a Larmor frequency measurement can then be accomplished by a measurement of the axial trap frequency. For single (anti)protons, the large mass of the particles necessitates the application of much stronger field curvatures than for the electron, because the continuous Stern-Gerlach effect is proportional to $\mu/m$. These field inhomogeneities are detrimental to a clean Larmor frequency measurement through a spatially varying resonance condition. Current efforts for the proton and antiproton therefore spatially separate the region of strong magnetic field curvature for spin state detection from a homogeneous magnetic field region, where spin transitions are driven and where the cyclotron frequency is measured (`double Penning-trap technique'\cite{haffner_double_2003}). Recently, single-proton single-spin-flip detection has been achieved\cite{mooser_resolution_2013}, a first direct measurement on antiprotons has been realized\cite{atrap_collaboration_one-particle_2013}, and the double Penning-trap technique has been demonstrated experimentally for protons\cite{mooser_demonstration_2013}. Still, due to the difficulties outlined above, experimental cycle times for the unambiguous detection of single-proton spin flips exceed an hour, limited by the typical time constants (100~s) of resistive cyclotron cooling used in the state of the art experiments\cite{mooser_resolution_2013}. The coupling of the cyclotron motion's magnetic moment to the magnetic field curvature leads to fluctuations in the measured axial frequency $\nu_z$, where the stability of $\nu_z$ scales with the square root of the average cyclotron quantum number $n_+$. In addition, differences in motional states and observed magnetic fields between the cyclotron and Larmor frequency resonance measurement may ultimately limit the accuracy. It is therefore highly desirable to ground-state-cool the cyclotron mode with highest efficiency, and to find alternative methods for spin state detection.

\section{Quantum logic approach}

Already in 1990, Heinzen and Wineland\cite{heinzen_quantum-limited_1990} proposed to measure $g$-factors of subatomic particles through their coupling to an atomic ion as a means of cooling the particle and detecting its spin state. The 1990 paper\cite{heinzen_quantum-limited_1990} envisaged storing the two particles in separate traps, and coupling their motional degrees of freedom through the image charges induced in a shared trap electrode. In 1998\cite{wineland_experimental_1998}, a direct coupling in a double-well potential was considered. Trapping the particles in spatially separate wells, coupled either directly or through a shared electrode, prevents a positively charged atomic ion and a negatively charged antiproton from annihilating. For the required spin-motional coupling, sideband transitions on the (anti)proton's spin-flip transition were considered, enabled by close proximity to a conductor (`near field').  

The basic ingredients of this scheme have recently been demonstrated at NIST with two atomic ions, rather than with an atomic ion and an (anti) proton. In a first experiment\cite{brown_coupled_2011}, the motion of two $^9$Be$^+$ ions in separate wells of a double-well potential was coupled at the single quantum level via the Coulomb interaction. In a second experiment\cite{ospelkaus_microwave_2011}, microwave near-fields were used to drive motional sidebands on a hyperfine transition in $^{25}$Mg$^+$, allowing the implementation of an entangling quantum logic gate. In the context of quantum information processing, compared to the typical laser-based quantum logic gates, this has important advantages in terms of integration, scalability, ease of control and projected fidelities. 

\begin{figure}
\centering
\includegraphics[width=0.8\textwidth]{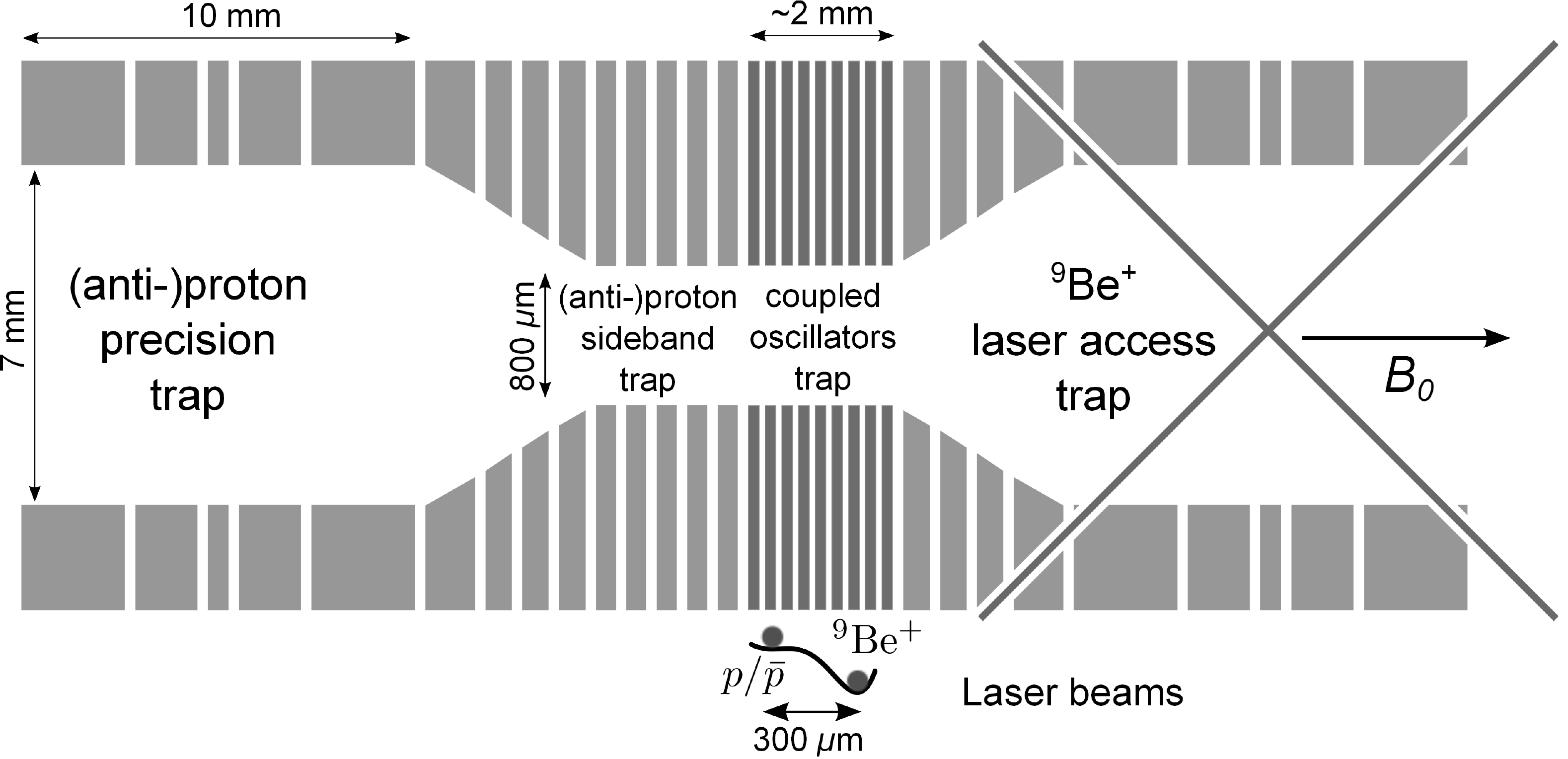}
\caption{Conceptual picture of Penning trap array for quantum logic readout and cooling of single (anti)protons (Dimensions not to scale)}
\label{aba:fig1}
\end{figure}

We are currently planning and constructing a setup based on the proposal by Heinzen and Wineland\cite{heinzen_quantum-limited_1990,wineland_experimental_1998} and on the above NIST experiments to realize a quantum logic based $g$-factor measurement apparatus for single (anti)protons. Our setup is based on a cylindrically symmetric Penning trap array (see Fig.~\ref{aba:fig1}). The different steps in the quantum logic cooling and readout procedure are carried out in spatially separate zones of the trap, a technique which has been used extensively in ion trap quantum computing with Paul traps. Our preliminary design features a narrow (800\,$\mu$m diameter) section of the trap with a double-well potential for a proton and a $^9$Be$^+$ ion separated by 300\,$\mu$m and a zone for spin-motional sideband transfer pulses for the (anti)proton. Our design also features sections with a large diameter for driving spin-flip transitions on single (anti)protons and for laser access to $^9$Be$^+$. Key advantages of the approach are: the projected short single-proton single spin-flip detection time of significantly less than 1\,s (typically exceeding one hour for the classical techniques) with important consequences in the presence of magnetic-field drifts; the ability to ground-state cool, with a corresponding reduction in magnetic-field-inhomogeneity induced uncertainties; finally the ability to use the atomic ion's Larmor frequency as a magnetic-field dependent flywheel in a $g$-factor comparison independent of the measurement of motional frequncies. The setup is currently under construction at Hannover. An initial demonstration of the approach will be done with protons. If successful, the laser systems and trap can be installed at the beamline of the BASE Collaboration, which is currently setting up an antiproton $g$-factor experiment based on the classical techniques and the double Penning trap technique at CERN. 

\section*{Acknowledgments}

We acknowledge funding by QUEST, LUH, PTB and RIKEN.


\begin{thebibliography}{xx}

\bibitem{bluhm_testing_1997}
R.\ Bluhm, V.A.\ Kosteleck\'y, and N.\ Russell, 
Phys.\ Rev.\ Lett.\ {\bf 79}, 1432 (1997).

\bibitem{bluhm_CPT_1998}
R.\ Bluhm, V.A.\ Kosteleck\'y, and N.\ Russell, 
Phys.\ Rev.\ D\ {\bf 57}, 3932 (1998).

\bibitem{dehmelt_csge}
H.\ Dehmelt and P.\ Ekstrom, 
Bull.\ Am.\ Phys.\ Soc.\ {\bf 18}, 72 (1973).

\bibitem{gabrielse_why_2009}
G.\ Gabrielse, 
Phys.\ Rev.\ Lett.\ {\bf 102}, 172501 (2009).

\bibitem{haffner_double_2003}
H.\ H\"affner \etal,
Eur.\ Phys.\ J.\ D {\bf 22}, 163 (2003).

\bibitem{mooser_resolution_2013}
A.\ Mooser \etal,  
Phys.\ Rev.\ Lett.\ {\bf 110}, 140405 (2013).

\bibitem{atrap_collaboration_one-particle_2013}
J.\ DiSciacca \etal,
Phys.\ Rev.\ Lett.\ {\bf 110}, 130801 (2013).

\bibitem{mooser_demonstration_2013}
A.\ Mooser \etal,
Phys.\ Rev.\ Lett.\ {\bf 723}, 78 (2013).

\bibitem{heinzen_quantum-limited_1990}
D.J.\ Heinzen and D.J.\ Wineland, 
Phys.\ Rev.\ A {\bf 42}, 2977 (1990).

\bibitem{wineland_experimental_1998}
D.J.\ Wineland \etal,
J.\ Res.\ Natl.\ Inst.\ Stand.\ Technol.\ {\bf 103}, 259 (1998).

\bibitem{brown_coupled_2011}
K.R.\ Brown \etal,
Nature {\bf 471}, 196 (2011).

\bibitem{ospelkaus_microwave_2011}
C.\ Ospelkaus \etal, 
Nature {\bf 476}, 181 (2011).

\end{thebibliography}
\end{document}